\documentclass[12pt,a4paper]{iopart}
\usepackage{iopams}  
\usepackage{graphicx}
\usepackage{color}
\usepackage{subcaption}
\usepackage[pdfborder={0 0 0}]{hyperref}
\usepackage{mciteplus}

\newcommand{\cbr}[1]{\left\{ #1\right\}}
\newcommand{\text}[1]{{\rm #1}}

\begin{document}

\title{A practical guide to event generation for prompt photon production with Sherpa}

\author{Frank Siegert}

\address{Institut f{\"u}r Kern- und Teilchenphysik, TU Dresden, D--01062 Dresden, Germany}
\ead{frank.siegert@cern.ch}

\begin{abstract}

The production of prompt photons is one of the most relevant scattering
processes studied at hadron-hadron colliders in recent years. This article will
give an overview of the different approaches used to simulate prompt photon
production in the Sherpa event generator framework. Special emphasis is placed
on a complete simulation of this process including fragmentation configurations.
As a novel application a merged simulation of
$\gamma\gamma$ and $\gamma\gamma+$jet production at NLO accuracy is presented
and compared to measurements from the ATLAS experiment.
\end{abstract}

%
%
%
%
%

\section{Introduction}

Prompt photon pairs have constituted one of the most relevant signatures in collider
measurements in the recent years. Not only the discovery of the Higgs
boson~\cite{Aad:2012tfa,*Chatrchyan:2012ufa}
was first announced in that channel, but also several further
searches for unknown heavy particles decaying into photons have been
performed by the LHC experiments~\cite{ATLAS-CONF-2016-059,*Khachatryan:2016yec}
and tracked closely by the theory community.

Though clearly the spotlight is on resonant production, one crucial
ingredient for such a physics programme to succeed has been the understanding of the
backgrounds from continuum diphoton production. In the direct search for
resonances this is often implemented by a fit to data from the invariant mass
distribution using the sideband regions next to the region of interest,
e.g.\ using various polynomial
functions~\cite{ATLAS-CONF-2016-059,*Khachatryan:2016yec}.

This is problematic as soon as one is looking for \textit{heavy}
resonances. Here, the fit of the functional parameters becomes unstable
because the high-mass sideband region contains only a low number of events
in data. Thus, it becomes crucial to have a theory-based prediction,
which will constrain the template fit of the Standard Model background
more strongly.

Despite its relevance, diphoton production has not been a very popular
testbed for the development and application of new and precise event
generation techniques up to now. This is partially due to its narrow scope
of application as a background process in searches for new particles, but to
a large extent has to be blamed on the intricate difficulties of prompt photon
production.

For example, the simultaneous production of photons and partons in any
higher-order calculation induces additional collinear divergences, which have
to be regularised and/or factorised into a fragmentation component. Furthermore,
the loop-induced gluon-gluon initial-state process, even though formally
only relevant at higher orders, is enhanced by the high gluon density in
the proton at LHC energies. Last, but not least, the calculation of
amplitudes with many photons and partons is challenging, which is
demonstrated by the slower progress in calculating NLO high-multiplicity
amplitudes for $\gamma\gamma$+jets~\cite{Badger:2013ava} as compared to
e.g.\ $V$+jets~\cite{Bern:2013gka}.

Theoretical predictions for prompt photon production are usually based on
fixed-order perturbation theory, assuming the application of
suitable isolation criteria~\cite{Glover:1993xc,*Catani:2013oma,Frixione:1998jh}
to remove infrared singularities. Additionally, a resummation of the large
logarithms induced by QCD corrections can be described semi-analytically by
photon fragmentation functions~\cite{LlewellynSmith:1978dc}.
They are conceptually similar to a parton distribution
function~ but describe the probability for the transition of a parton
into a collinear photon with a given momentum fraction.

Perturbative calculations of higher-order QCD corrections to single photon and
diphoton production were performed in the
JETPHOX~\cite{Catani:2002ny,*Aurenche:2006vj,*Belghobsi:2009hx} and
DIPHOX~\cite{Binoth:1999qq,*Binoth:2000zt} programs at NLO,
for the first time including also the full fragmentation contribution.
Furthermore, in recent years also NNLO accurate results for diphoton
production have appeared~\cite{Catani:2011qz,*Campbell:2016yrh} and yield an
improved agreement with experimental
measurements~\cite{Aad:2012aa,Aaltonen:2012jd,*Abazov:2013pua,*Chatrchyan:2014fsa}.

For usage in experimental measurements and direct comparisons to those, it is
necessary to go beyond analytical calculations and include hadronisation and
underlying-event effects in the prediction. These are typically generated by
parton shower event generators, which not only include a phenomenological
modelling of hadronisation and multiple particle interactions but also add
to the fixed-order calculation a resummation of large logarithms due to
collinear parton emissions, thus providing an alternative to the usage of
analytical fragmentation functions.
Several modern Monte Carlo event
generators~\cite{Hoeche:2009xc,Odaka:2016tef,D'Errico:2011sd,Jezo:2016ypn}
are going beyond the simple parton-shower based approach and
allow for a more precise simulation of prompt photon processes.

This article consists of two parts: Section~\ref{sec:review} contains a
pedagogical overview of the traditional and modern methods for the simulation
of photon production in the hadron collider event generator Sherpa.
In Section~\ref{sec:mepsnlo} the Sherpa framework is then used to introduce
for the first time a merged simulation of $\gamma\gamma$ and $\gamma\gamma$+jet
at NLO QCD accuracy. After presenting the results obtained with this simulation
the article will conclude and give an outlook towards further work.

\section{Overview of prompt photon event generation in Sherpa}
\label{sec:review}

This section contains a short overview of how prompt photon production is
simulated in full event generators like Sherpa. It starts from the basic
parton shower picture, and then describes the more advanced approaches.
While the description is mostly kept general, the details and examples will
refer to the implementation in the Sherpa event
generator~\cite{Gleisberg:2008ta}.

\subsection{Basic parton shower}
\label{sec:ps}

\begin{figure}[tbp]
  \centering
  \begin{subfigure}[b]{.3\linewidth}
    \centering
    \includegraphics[width=0.8\textwidth]{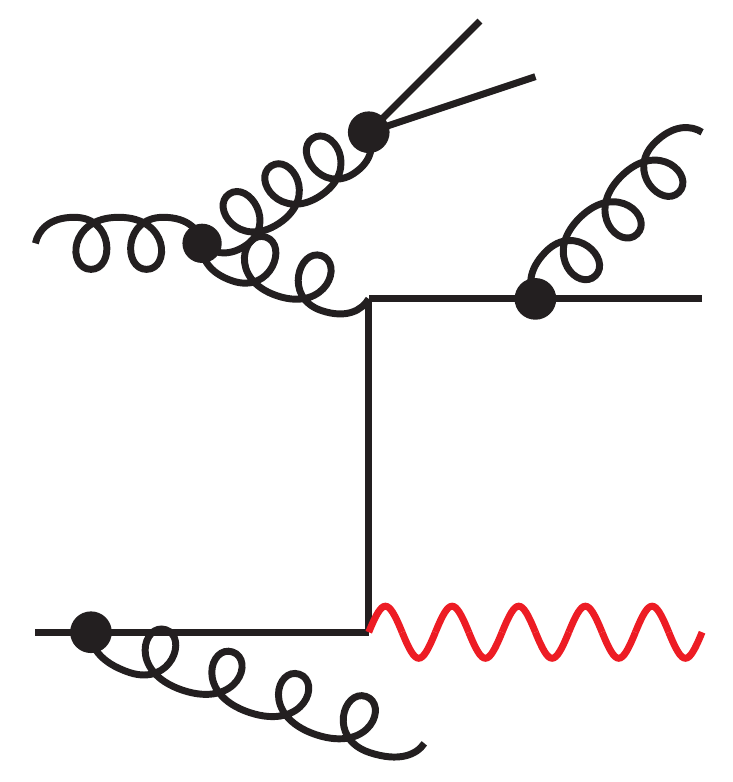} 
    \caption{Direct photons}
    \label{fig:psdirect}
  \end{subfigure}%
  \begin{subfigure}[b]{.3\linewidth}
    \centering
    \includegraphics[width=0.8\textwidth]{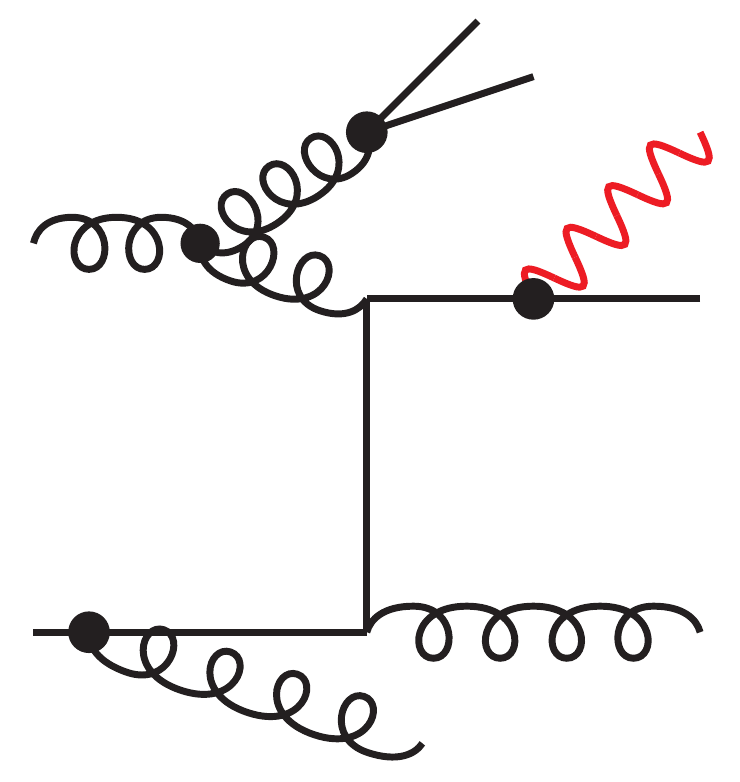} 
    \caption{Fragmentation photons}
    \label{fig:psfrag}
  \end{subfigure}%
  \begin{subfigure}[b]{.4\linewidth} 
    \centering
    \includegraphics[width=0.8\textwidth]{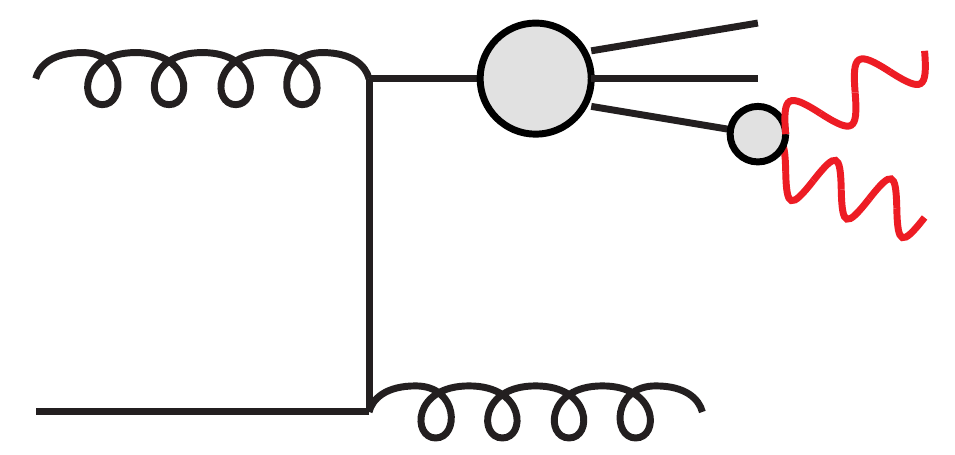} 
    \vspace*{6mm}
    \caption{Non-prompt photons}
    \label{fig:psnonprompt}
  \end{subfigure}
  \caption{Photon production mechanisms in traditional parton shower programs. The black dots represent parton shower splittings, while the grey circles represent hadronisation and hadron decay processes.}
\end{figure}

In a traditional parton shower simulation, based on $2\to 2$ matrix elements
and a subsequent parton cascade, photons can be produced by three different
mechanisms:
\begin{enumerate}
\item Direct production

  Matrix elements for $pp \to \gamma +X$ production are calculated at the
  leading order and the parton shower adds QCD emissions,
  cf. Fig.~\ref{fig:psdirect}.

  This implies a
  strict hierarchy of scales in the shower evolution variable $t$, with QCD
  parton emissions $t_{\rm QCD}$ allowed only at a lower (softer) scale than the
  factorisation scale defined by the direct photon production process
  $t_{\rm QED}$.

\item Fragmentation production

  Parton showers can be extended straightforwardly to include not only QCD
  splitting functions, but also their QED
  equivalent~\cite{Seymour:1991xa,*Seymour:1994bx}. Thus, photons
  can be emitted from quark lines, resumming the corresponding logarithmic
  enhancement from collinear configurations.

  This implies a combined shower evolution with a ``competition'' for the
  emission phase space between QED and QCD emissions. Since QCD and QED
  emissions do not interfere, the Sudakov form factor takes a factorised
  form~\cite{Hoeche:2009xc}:
  \begin{equation}\label{eq:sudakov_fac}
    \Delta(\mu_0^2,Q^2)\,=\;\Delta^{\rm QCD}(\mu_0^2,Q^2)\;\Delta^{\rm QED}(\mu_0^2,Q^2)\;,
  \end{equation}
  where the QED part,
  \begin{equation}\label{eq:sudakov_qed}
    \Delta^{\rm QED}(\mu_0^2,Q^2)\,=\;
    \exp\cbr{\,-\int_{\mu_0^2}^{Q^2}\frac{{\rm d} t}{t}
      \int{\rm d} z\,\sum\limits_{i}
      \frac{1}{2}\,{\mathcal{K}}_{i}^{\,\rm QED}(z,t)\;}\;,
  \end{equation}
  contains the QED splitting functions ${\mathcal{K}}_{i}^{\,\rm QED}(z,t)$,
  in direct analogy to the QCD part.

  The production of photon pairs
  through this fragmentation mechanism requires the inclusion of
  matrix elements for $pp \to \gamma \textrm{ + jet}$ (single fragmentation)
  and even $pp \to \textrm{jet + jet}$ (double fragmentation) in the
  simulation. A pictorial representation is shown in Fig.~\ref{fig:psfrag}.

  Again, as this is a parton shower, one encounters a strict hierarchy of
  scales opposite to the case above, with the QED emissions $t_{\rm QED}$ being
  softer than the core parton production process $t_{\rm QCD}$.

\item Non-prompt production

  Hadrons decaying into photons, like $\pi^0 \to \gamma\gamma$,
  see Fig.~\ref{fig:psnonprompt}, represent
  a photon contribution that is always present, but can be disentangled
  from the prompt production above in a physical and to some
  extent even experimentally meaningful way, and shall thus not be discussed
  in the course of this work.
\end{enumerate}

\begin{figure}[tbp]
  \centering
  \includegraphics[width=0.48\textwidth]{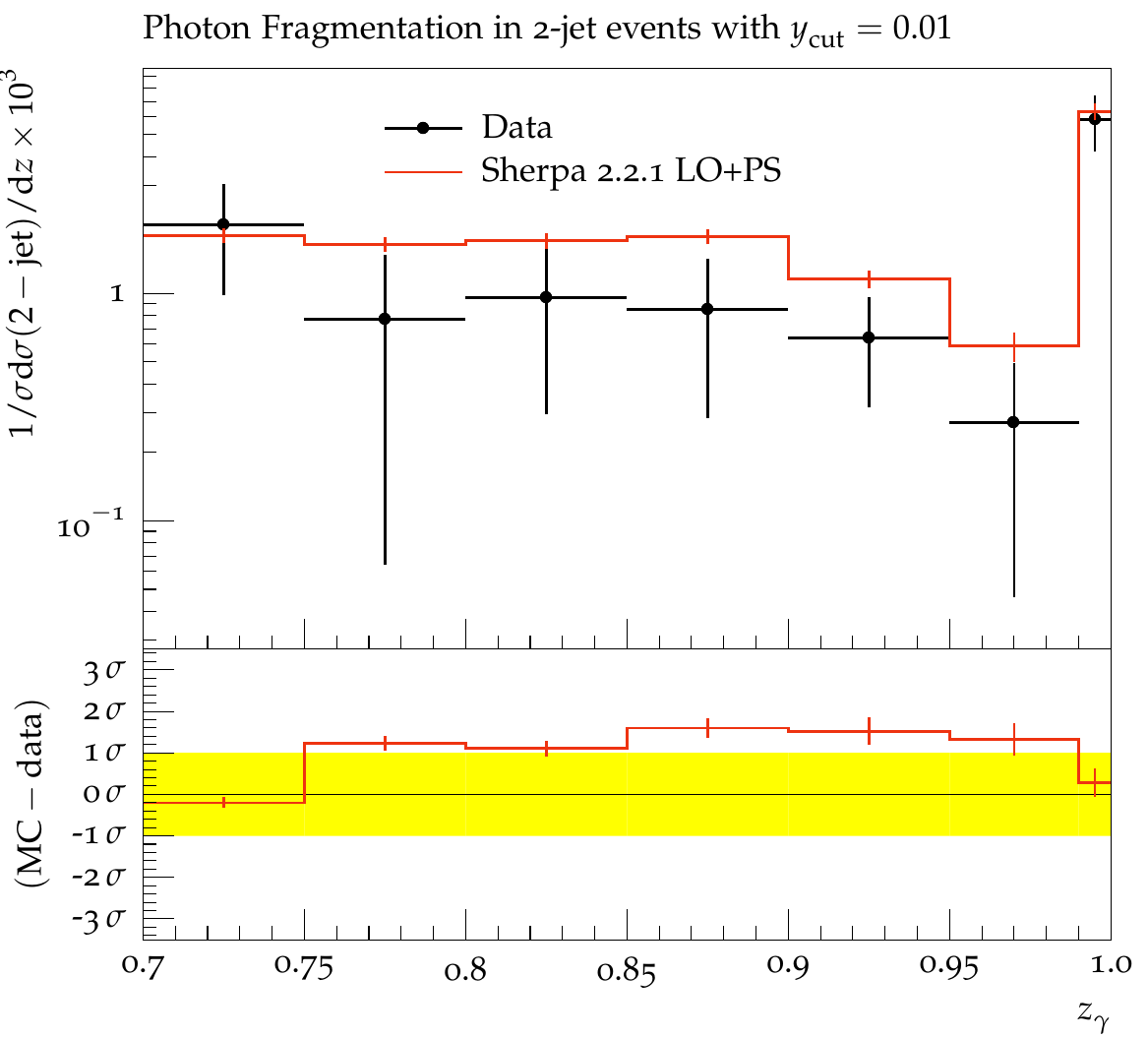}\hfill 
  \includegraphics[width=0.48\textwidth]{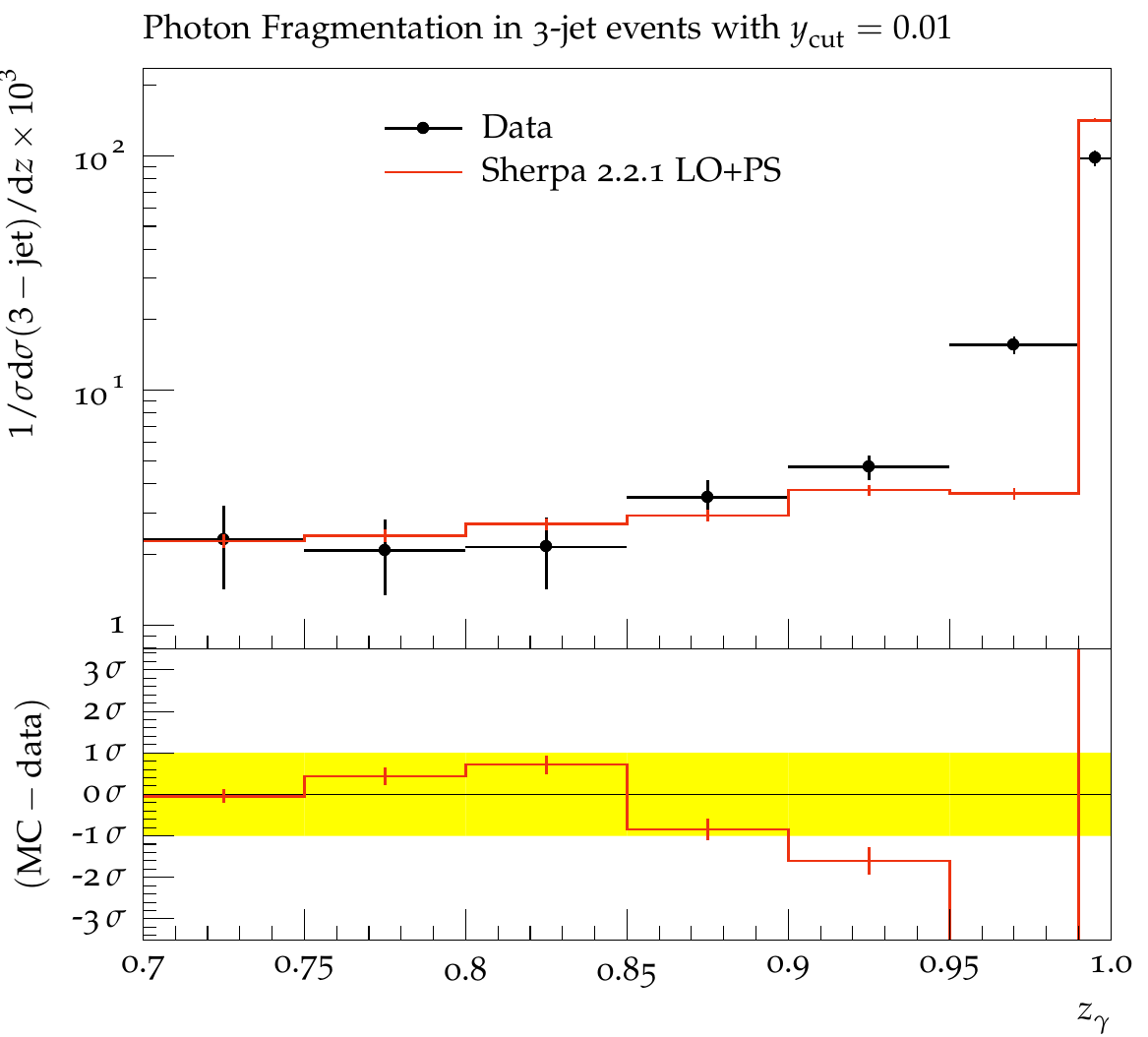} 
  \caption{Fragmentation distributions measured by the ALEPH experiment in comparison with a QED shower prediction from Sherpa 2.2.1.}
  \label{fig:aleph}
\end{figure}

\noindent The fragmentation production in a parton shower can be probed
very specifically by looking at photons in jets produced in electron-positron
collisions. To demonstrate the QED shower implementation in Sherpa
2.2.1~\cite{Hoeche:2009xc}, it is compared to the measurement of fragmentation
distributions by the ALEPH experiment~\cite{Buskulic:1995au} in
Fig.~\ref{fig:aleph}.
Those are differential in the energy fraction the photon carries within the
jet, $z_\gamma$, and shown for 2- and 3-jet events.
While the experimental uncertainties are very large, and the agreement is not
perfect in all bins, the general features of the fragmentation distribution
are described fairly well already by this simple parton shower prediction.

It should be noted, that direct and fragmentation production cannot
be separated in a physical way. The division into these two parts is
merely a technical detail of the calculation or simulation, and they cannot
be separately measured.

There are two significant drawbacks of the pure parton shower approach.

From a purely practical point of view, the prediction of the fragmentation
component can be tedious, as the combination of a low probability for
the emission of a hard and isolated photon with the high cross section of
the underlying $2\to 2$ processes requires the generation of incredibly
large event samples. This can be mitigated to some extent by using shower
programs where it is possible to enhance the QED splitting processes using
appropriately weighted events~\cite{Hoeche:2009xc}.

But more importantly, the accuracy of the simulation is at best leading order
(for the direct component) or even takes into account only the leading
logarithms through the parton shower resummation (for the fragmentation
component).

Nevertheless, these methods allow to reach a fair agreement with prompt
photon production data from
experiments~\cite{Aad:2012aa,Aaltonen:2012jd,*Abazov:2013pua,*Chatrchyan:2014fsa}.

\subsection{Tree-level QCD merging}
\label{sec:qcdmeps}

Parton-shower simulations can be improved by the inclusion of tree-level QCD
multi-jet matrix elements in a consistent merging
scheme~\cite{Catani:2001cc,*Lonnblad:2001iq,*Mangano:2001xp,*Krauss:2002up,*Hamilton:2009ne,*Lonnblad:2011xx,Hoeche:2009rj}.
Such simulations have become the workhorses of LHC experiments for the
simulation of final states including multiple jets and can be applied to prompt
photon final states as well.

The main idea is the introduction of a separation criterion $Q_{ij}$ for
partons $i, j$ which is then used to slice the phase space for parton
emissions from a given core process $pp\to X$ into two domains through
a merging cut $Q_{\rm cut}$:

\begin{description}
\item[ME domain] $Q_{ij}>Q_{\rm cut}$

  This region is populated by the tree-level matrix element including an
  additional parton, $pp\to X \textrm{ + jet}$.
\item[PS domain] $Q_{ij}<Q_{\rm cut}$

  This region is filled using the core matrix element $pp\to X$ and
  performing a parton shower emission.
\end{description}

\begin{figure}[tbp]
  \centering
  \includegraphics[width=0.9\textwidth]{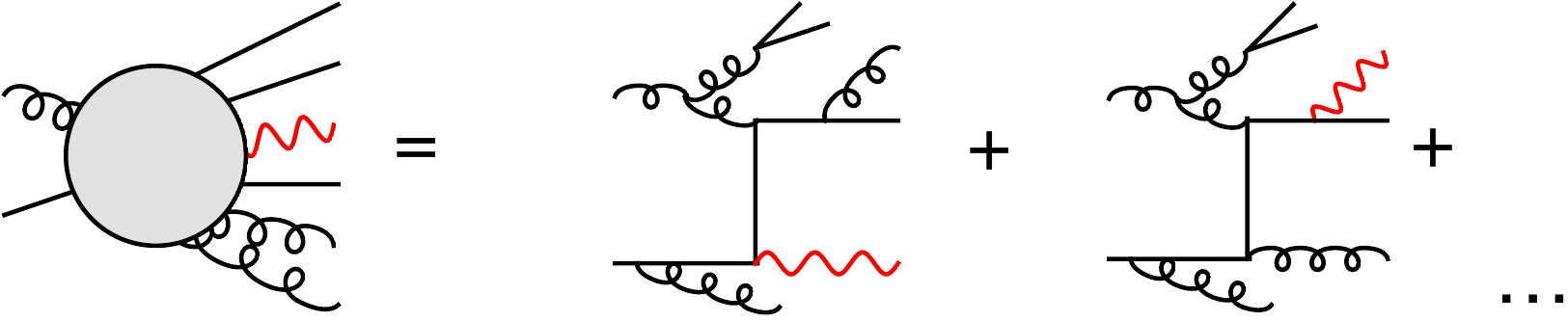} 
  \caption{Example for multi-jet process within QCD-merged setup, demonstrating the inclusion of the contributions from Figs.\ \ref{fig:psdirect} and \ref{fig:psfrag}.}
  \label{fig:qcdmeps}
\end{figure}

It becomes particularly interesting when this is applied to prompt photon
production, depicted in Fig.~\ref{fig:qcdmeps}.
There is no longer the immediate correspondence of ``direct $\equiv$ ME''
 and ``fragmentation $\equiv$ PS'' production.
Instead, the multi-jet matrix elements will now also contribute to the phase
space which was previously covered by the QED parton shower.

Trying to relate these to each other, it is instructive to consider the
scale hierarchies mentioned in Sec.~\ref{sec:ps}. For example,
considering the first emission, the corresponding matrix element covers the
phase space including both hierarchies:

\begin{equation}
  Q_{ij}>Q_{\rm cut} \quad\cup\quad t_{\rm QCD}<t_{\rm QED},
\end{equation}
\begin{equation}
  \label{eq:invhier}
  Q_{ij}>Q_{\rm cut} \quad\cup\quad t_{\rm QCD}>t_{\rm QED}.
\end{equation}

The QED shower is disabled in such a scheme, and all photon emissions will
be generated directly with matrix-element accuracy.
However, the replacement of the QED shower by higher-order
tree-level matrix elements also comes with disadvantages.

The inclusion of high
multiplicity matrix elements in the QCD merging approach makes it necessary
to manually impose a regularisation
of the infrared singularities of the photon in multi-parton configurations.
To this end, phase space cuts are applied, which mimick experimental
requirements for the transverse momentum of the photon and its isolation
with respect to hadronic activity. The event generation is thus not fully
inclusive anymore with respect to the photon, and care has to be taken to
not introduce a bias for example for events after detector simulation.

Depending on how inclusive these photon cuts are chosen, it can also
become problematic, that the resummation of collinear QED emission terms is not
included anymore in the QCD merged sample. But in practice, this will be
mitigated by the typical photon energy and isolation cuts in any prompt photon
analysis.

\subsection{Tree-level QCD \texorpdfstring{$\otimes$}{x} QED merging}
\label{sec:qcdqedmeps}

A natural extension of the QCD ME+PS merging approach to QED emissions
leads to a solution of both problems just mentioned. In a combined
QCD $\otimes$ QED merging, one incorporates QED emissions from both the
parton shower and the hard scattering matrix elements. To avoid the
resulting double-counting, the parton separation criterion $Q_{ij}$ is extended
also to photons, effectively imposing a democratic merging of photons and
partons.

\begin{figure}[tbp]
  \centering

  \includegraphics[width=0.48\textwidth]{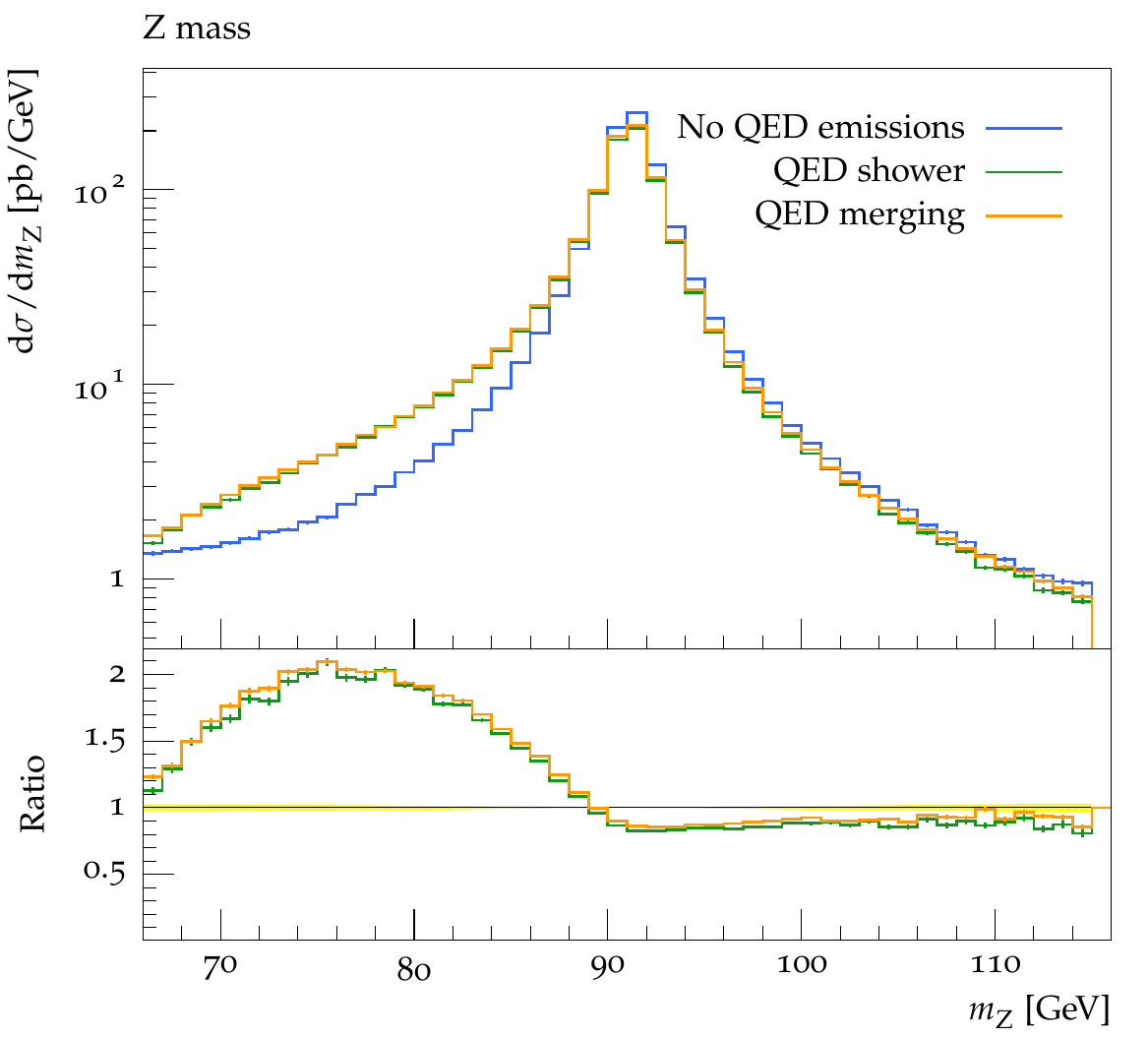}\hfill 
  \includegraphics[width=0.48\textwidth]{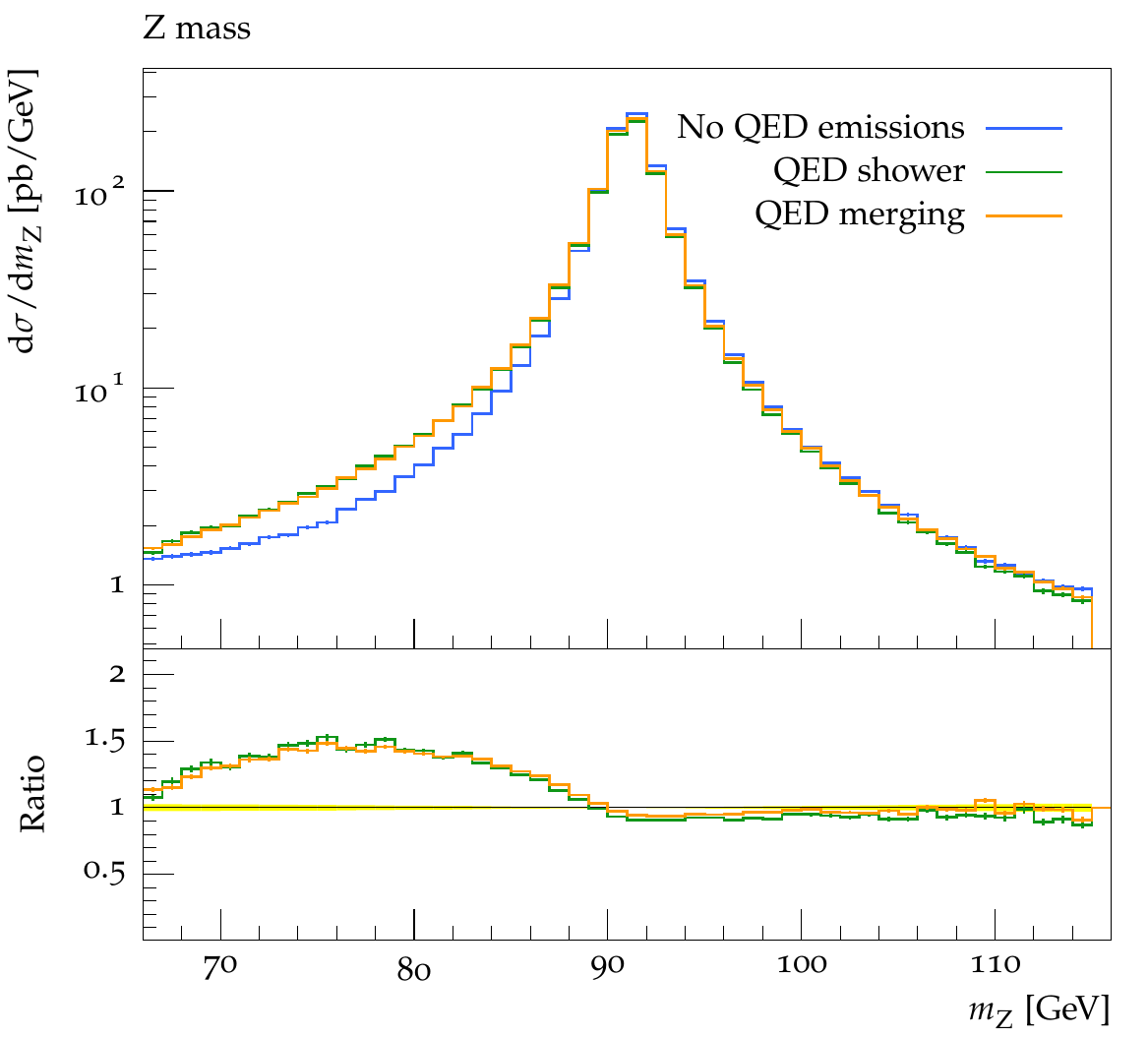} 
  
  \caption{QED merging validation in the $Z$-lineshape in lepton-pair production at the 13 TeV LHC, using bare leptons (left) and dressed leptons (right).}
  \label{fig:qedmepsmll}
\end{figure}

To demonstrate first the QED merging component alone,
lepton pair production at the 13 TeV LHC is chosen as a testbed.
The benchmark observable in Fig.~\ref{fig:qedmepsmll} is the $Z$-lineshape
which is sensitive to QED emissions, in particular when built from ``bare''
leptons (left) as opposed to ``dressed'' leptons (right), which are built by
including photon radiation in a cone of $R=0.1$ around the lepton.
The QED merging is here tested using an extremely low merging scale,
$Q_{\rm cut}=0.1\textrm{ GeV}$, to have any effect on this observable at all.
Matrix elements for the $pp\to \ell\ell$ production process with
up to two additional photons are included in the merged sample.

The results from this exercise demonstrate that such a merging approach
could solve both disadvantages mentioned in the last section:

The merged prediction reproduces the shower prediction for the lineshape,
i.e.\ the resummation of QED emissions is implemented
correctly in this approach, solving the first problem mentioned earlier.

Furthermore, also the inclusiveness of the event sample with respect to
any potential photon cuts is restored: the cuts applied to the higher-multiplicity
matrix elements now become \emph{merging scales}, which only govern the
transition to the parton shower emissions, that fill the phase space.

The combined QCD $\otimes$ QED merging can now be implemented
in a straightforward way and can thus be applied to prompt photon production
by combining matrix elements for multi-parton and multi-photon production
completely democratically.

At this point the reader may argue that one is impacted again by the
inefficiencies of the QED parton shower discussed in Sec.~\ref{sec:ps}.
In other words, the relative cross section of the tree-level matrix
elements containing parton production will be orders of magnitude larger
than the matrix elements with outgoing photons -- but both appear within the
same merged setup.
While this large cross section difference is, of course, compensated by the
low probability to produce a hard isolated photon in the QED shower, it
remains impractical to generate such a sample for hard isolated photons.

This issue can be mitigated by choosing the merging cut for photon emissions
smartly, such that QED {\it shower} emissions are not relevant for the
analysis region.
One is at liberty to choose a different value of $Q_{\rm cut}$ for
QED emissions compared to QCD emissions.
This is reasonable because it reflects the different behaviour
of both the theory in the infrared limit and the experimental resolution.
Since the merging criterion is a technical parameter which merely has
to reproduce the proper singular limits, one can even choose a different
functional form of $Q_{ij}$ for the case where $i$ or $j$ is a photon. In
effect, one can adapt the merging cut to reflect the energy and isolation cuts
of hard isolated photons in the analysis. For example, in the common case where
a minimal transverse momentum and a spatial separation
$\Delta R_{ij}^2=\Delta\eta_{ij}^2+\Delta\phi_{ij}^2\geq D$ are required one
could choose the equivalent of the longitudinally invariant
$k_t$-algorithm~\cite{Catani:1992zp,*Catani:1993hr,Hoeche:2009xc}:
\begin{equation}\label{eq:jetcrit_photons}
  Q_{ij}^2\,=\;\min\cbr{p_{\perp,i}^2,p_{\perp,j}^2}\frac{\Delta R_{ij}^2}{D^2}
  \quad \textrm{and} \quad
  Q_{ib}^2\,=\;p_{\perp,i}^2\;,
\end{equation}
where $i$ is always a photon, $j$ is a charged final-state particle,
and $b$ represents a charged incoming particle.

Going even further, with an appropriately chosen merging cut, the contributions
due to the QED shower can be ignored in the final event generation sample, and
be generated merely as a separate low-statistics sample to cross-check that
they are indeed not contributing to the region of interest.

Instead of pursuing this route further, let us now turn to a new approach to
generate prompt photon production at higher accuracies.

\section{Merged diphoton production at NLO QCD accuracy}
\label{sec:mepsnlo}

The existing methods for hadron-level event generation described earlier all
have one significant drawback compared to state-of-the-art fixed-order
calculations: they are only accurate to leading order in QCD perturbation
theory.

Extending NLO accuracy to the full QCD $\otimes$ QED merging
setup as described in Sec.~\ref{sec:qcdqedmeps} would necessitate the inclusion of
higher order corrections not only in QCD but also in QED.
Some of the building blocks necessary for such a combined evolution at NLO
accuracy are already available, such as NLO QED matrix elements
including dedicated subtraction schemes for QED
emissions~\cite{Dittmaier:1999mb,*Dittmaier:2008md}, and
leading-order QED splitting functions as implemented in the parton showers
described in Sec.~\ref{sec:ps}.
But several crucial ingredients for a NLO QCD $\otimes$ QED merging are missing,
such as the matching of QED real emission matrix elements with the QED parton
shower, as well as the implementation of the subtracted truncated shower veto
in the QED part of the MEPS@NLO merging~\cite{Hoeche:2012yf}.

One could even go one step further, beyond the factorised combination of
QCD and QED corrections, and aim for a simulation with the full combined
$\mathcal{O}(\alpha_s \alpha)$ corrections. To that end, the parton shower
would have to be extended to be fit for NLO splitting functions and
contain kernels up to
$\mathcal{O}(\alpha_s \alpha)$~\cite{deFlorian:2015ujt}.

Here, an intermediate step is presented, allowing to achieve NLO QCD accuracy
as an extension of the tree-level QCD merging approach described above.
While there has been previous work to match NLO
matrix elements for the \emph{inclusive} diphoton production process with a parton
shower~\cite{D'Errico:2011sd,Jezo:2016ypn}, this article goes one step further and 
includes NLO accuracy for both $pp\to \gamma\gamma$ and $pp\to \gamma\gamma$+jet
production in a merged simulation.
This allows to benefit from the features introduced in
Sec.~\ref{sec:qcdmeps}, in particular the generation of the fragmentation
component with the help of higher-multiplicity matrix elements instead of
the QED shower.

\subsection{Programs and setup}
\label{sec:setup}

The studies shown in the following are done using the Sherpa event
generator framework~\cite{Gleisberg:2008ta}, including
a parton shower based on Catani-Seymour subtraction
terms~\cite{Schumann:2007mg}, tree-level matrix elements from the
Amegic~\cite{Krauss:2001iv} and Comix~\cite{Gleisberg:2008fv} generators
and one-loop matrix elements from OpenLoops~\cite{Cascioli:2011va}.
The default Sherpa 2.2 tuning is used together with the NNPDF 3.0 NNLO PDF
set~\cite{Ball:2014uwa}.

The matching of NLO QCD matrix elements with the parton shower is implemented
using a variant~\cite{Hoeche:2011fd} of the original MC@NLO
method~\cite{Frixione:2002ik}.

NLO+PS matched simulations for $pp\to \gamma\gamma$ and
$pp\to \gamma\gamma$+jet production are then merged into an inclusive sample
using the MEPS@NLO approach~\cite{Hoeche:2012yf}.
Additionally, matrix elements with up to three partons in the final state are
included at LO accuracy in the approach of~\cite{Hoeche:2010kg}.

The QCD factorisation and renormalisation scales for the diphoton core
process are set to the invariant mass of the photon pair. The QED coupling
constant is set to $\alpha=1/137.036$. Additional QCD couplings in the
higher-multiplicity matrix elements are evaluated at their reconstructed
branching scale in the formalism of~\cite{Hoeche:2009rj}.

The matrix elements are generated with cuts on the transverse momentum
of the leading (subleading) photon,
$p_\perp^{\gamma}>21 (18) \textrm{ GeV}$.
Photons are required to be central within $|y_\gamma|<2.7$ and isolated
at the parton level according to a smooth cone isolation~\cite{Frixione:1998jh}
with parameters $\delta=0.1$, $n=2$ and
$\epsilon=0.1$. Additionally, a separation cut between
both photons of $\Delta R(\gamma_1,\gamma_2)>0.2$ is applied.

\subsection{Dynamical merging scale}
\label{sec:dynqcut}

As NLO accurate QCD $\otimes$ QED merging
is not available in current event generators, one has to resort to the
QCD merging approach as described in Sec.~\ref{sec:qcdmeps}, but now at the NLO
accuracy level. To deal with the limitations of the QCD merging approach,
the following idea is implemented:

The main problem, namely the limited inclusiveness of the ``fragmentation''
component, concerns the inverse hierarchy configurations.
While the ME domain part, Eq.~(\ref{eq:invhier}), is included, the complementary
contribution stemming from a QCD parton shower emission,
\begin{equation}
  \label{eq:invhierps}
  Q_{ij}<Q_{\rm cut} \quad\cup\quad t_{\rm QCD}>t_{\rm QED}
\end{equation}
would be necessary to generate the full ``fragmentation'' component.
Since the parton shower cannot generate emissions with this inverse
hierarchy, this contribution remains uncovered.

At first sight, this does not appear to be a significant issue, since the
phase space described in Eq.~(\ref{eq:invhierps}) requires an emission which
is still in the PS domain but harder than the photon production scale, thus
naively making this contribution negligible.
But in the case of prompt photon production there is no hard (mass) scale in
the process which would automatically set a lower boundary for the shower
starting scale. Hence this phase space region can become significant.

To solve this problem, one can make use of an approach which has been pioneered
in a similar situation with low factorisation scales, namely for QCD merging
in DIS~\cite{Carli:2009cg}.

The basic idea is to introduce a dynamical merging cut $Q_{\rm cut}$ such
that it is smaller than the shower starting scale $t_{\rm QED}$ at each phase
space point.
The shower starting scale is usually given by the factorisation scale of the
prompt photon production process, i.e.\ something like $\mu_F=p_\perp^\gamma$
for single photon production or $\mu_F=m_{\gamma\gamma}$ for photon pair
production. The dynamical merging cut can then be built from a fixed value
$\bar{Q}_{\rm cut}$ and the factorisation scale $\mu_F$ as:

\begin{equation}
  \label{eq:dynqcut}
  Q_{\rm cut}=\left[ \frac{1}{\bar{Q}_{\rm cut}^2} + \frac{1}{S^2\,\mu_F^2}\right]^{-1/2}.
\end{equation}

This particular form has the advantage, that it provides a smooth
interpolation between the fixed value and the factorisation scale,
but is dominated by the smaller of the two. The safety factor $S \lesssim 1$
can be chosen according to the parton shower model.
The fixed component $\bar{Q}_{\rm cut}$ is present in this dynamical scale to
ensure that matrix element accuracy is used for hard emissions in phase space
regions with high $\mu_F$.

\subsection{Results}
\label{sec:results}

The approach described above is validated with a comparison to
measurements from the ATLAS experiment~\cite{Aad:2012aa} through their
implementation in Rivet~\cite{Buckley:2010ar}.

Events of proton-proton collisions at $\sqrt{s} = 7\textrm{ TeV}$ are selected
to contain two isolated photons with transverse energies above 25 GeV
and 22 GeV, both within $|y_\gamma|<1.37$ or $1.52<|y_\gamma|<2.37$ and with
$\Delta R(\gamma,\gamma)>0.4$. To define isolated photons, the hadronic
energy within a cone of $\Delta R<0.4$ but excluding the inner core
of $|\Delta \eta| < 0.0625$ and $|\Delta \Phi| < \frac{7\pi}{256}$ is
required to be $E_{\rm cone}<4\textrm{ GeV}$. The energy is corrected 
taking into account the median transverse energy density in this region on an
event-by-event basis to limit the sensitivity to multiple parton and proton
interactions.

Predictions are obtained both for the tree-level as well as NLO QCD merging
approach using the programs and setup described in Sec.~\ref{sec:setup}.
The dynamical merging scale introduced in Sec.~\ref{sec:dynqcut} is used
in both cases with $\bar{Q}_{\rm cut}=10$~GeV.
In addition to the central prediction an uncertainty band is estimated from a
7-point independent scale variation of the factorisation and renormalisation
scales by a factor of two up and down, excluding only opposite variations.
Furthermore central predictions from switching the PDF set in the matrix
element to CT14nnlo~\cite{Dulat:2015mca} and
MMHT2014nnlo~\cite{Harland-Lang:2014zoa} are displayed.
These variations are calculated using on-the-fly event weights in the
implementation of~\cite{Bothmann:2016nao}.

Comparing the LO and NLO predictions for the total fiducial cross section
in Table~\ref{tab:xs} one can draw two conclusions:
The MEPS@NLO prediction agrees much better with the measured value than the
MEPS@LO prediction. Both are compatible within uncertainties, but the size
of the uncertainties in the former is reduced significantly.

\begin{table}[tbp]
  \centering
  \begin{tabular}{lll}
    $\sigma_{\rm MEPS@LO}$ [pb] & $\sigma_{\rm MEPS@NLO}$ [pb] & $\sigma_{\rm ATLAS}$ [pb] \\
    \hline
    $33.9^{+9.6 (28\%)}_{-5.9 (18\%)}$ & $44.8^{+6.7 (15\%)}_{-6.5 (15\%)}$ & $44.0^{+3.2(7\%)}_{-4.2(10\%)}$ \\
  \end{tabular}
  \caption{Total fiducial cross sections and systematic uncertainties from MEPS@LO and MEPS@NLO predictions compared to the ATLAS measurement.}
  \label{tab:xs}
\end{table}

\begin{figure}[tbp]
  \centering
  \includegraphics[width=0.48\textwidth]{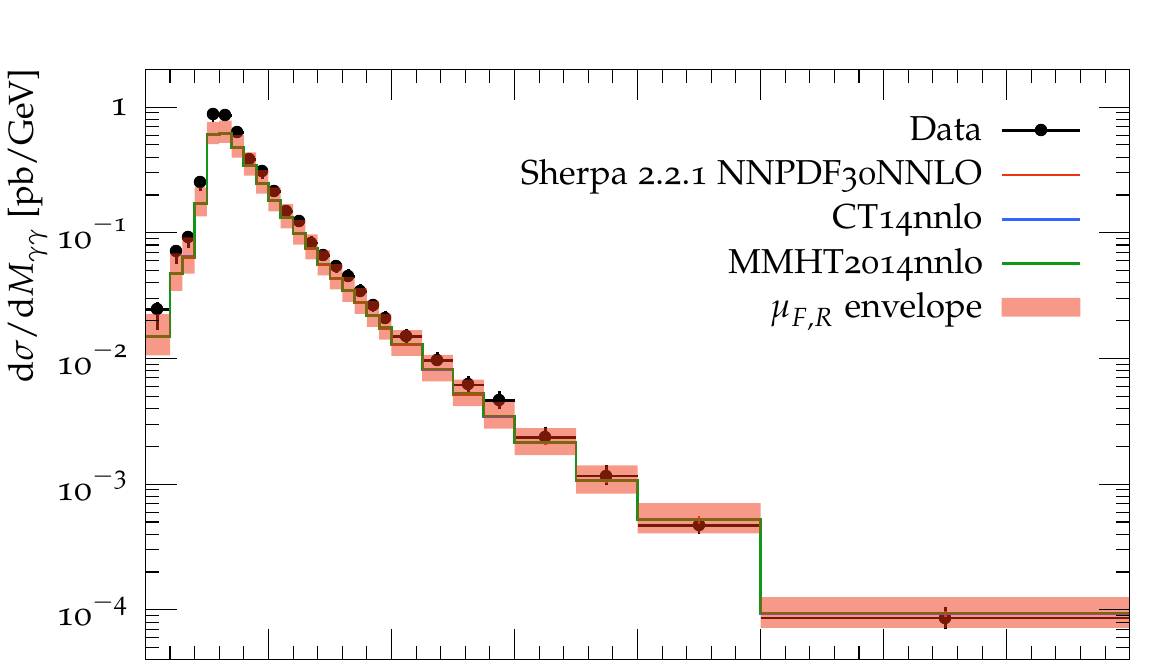}\hfill 
  \includegraphics[width=0.48\textwidth]{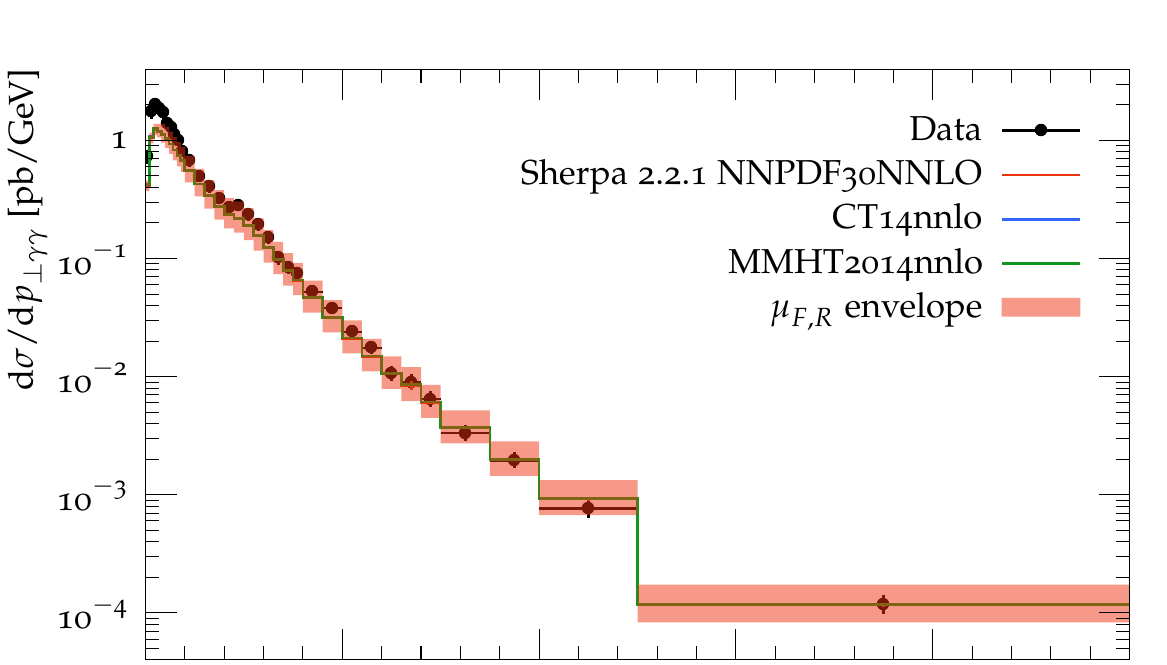}\\ 
  \includegraphics[width=0.48\textwidth]{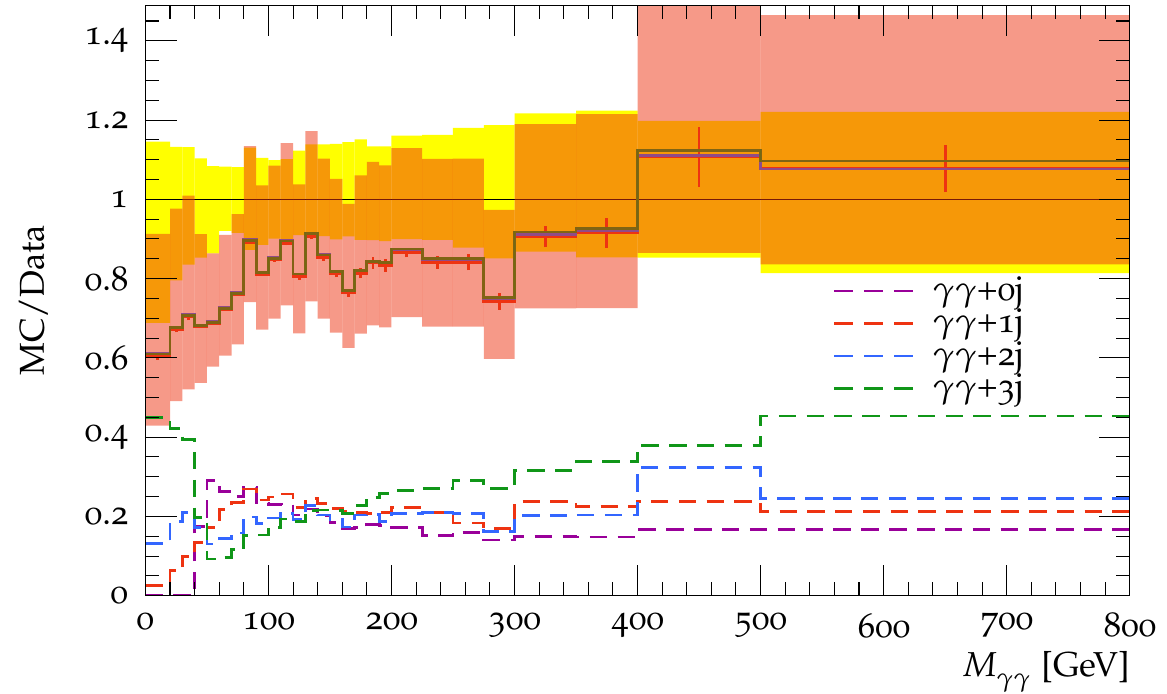}\hfill 
  \includegraphics[width=0.48\textwidth]{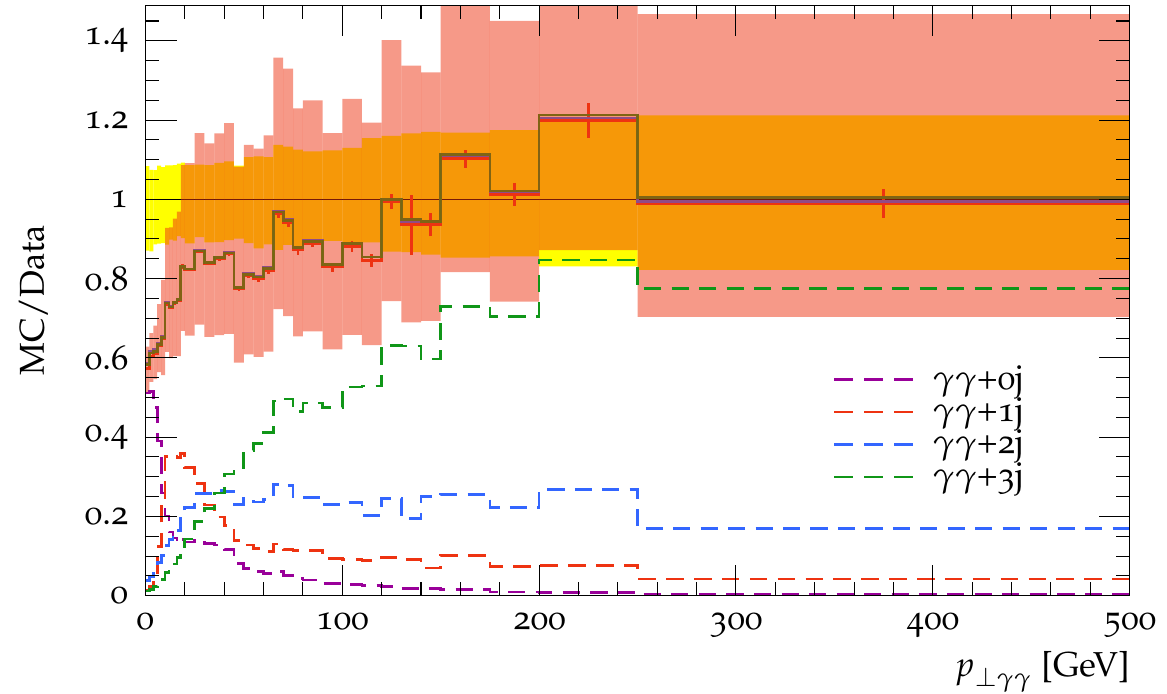}\\ 
  \includegraphics[width=0.48\textwidth]{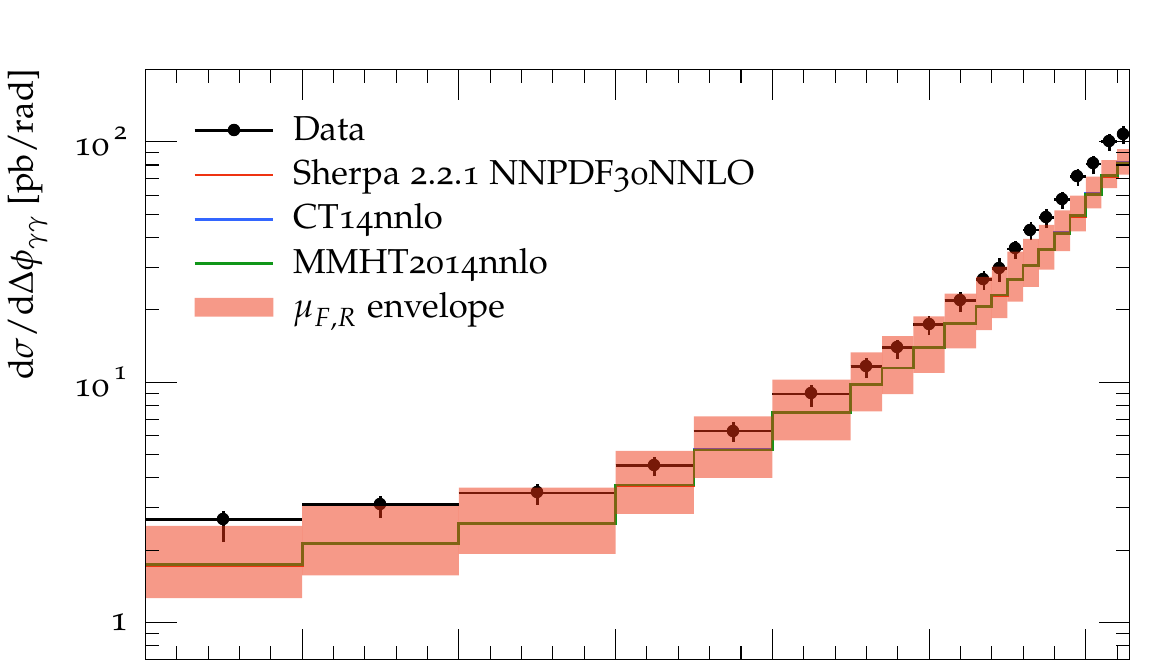}\hfill 
  \includegraphics[width=0.48\textwidth]{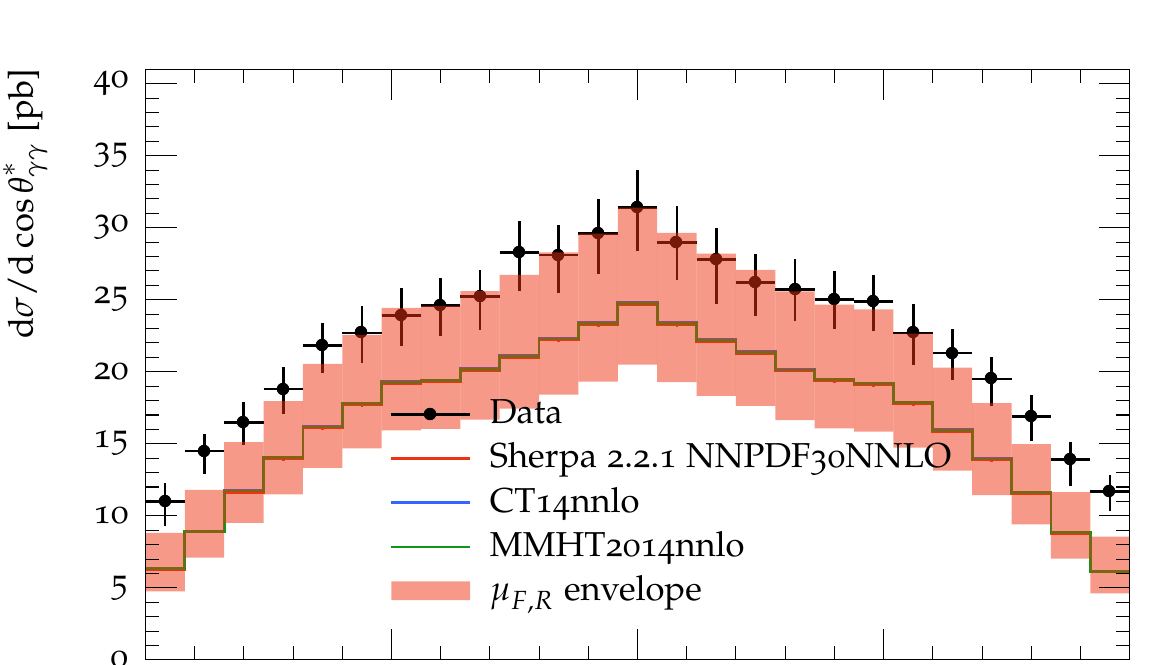}\\ 
  \includegraphics[width=0.48\textwidth]{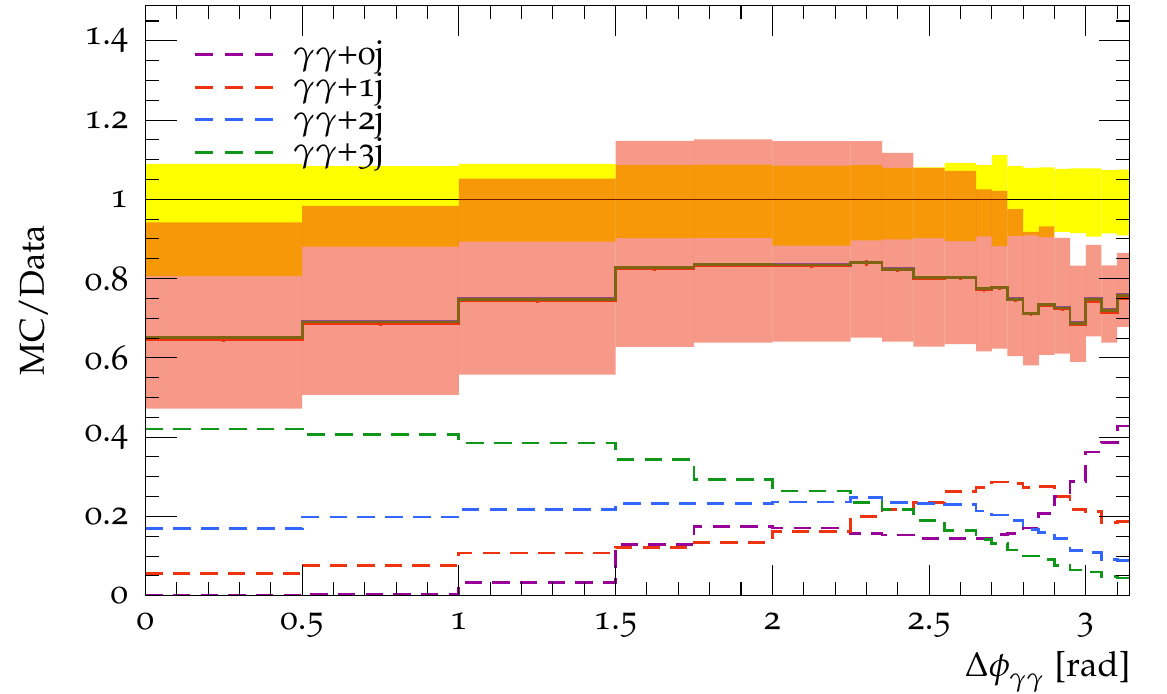}\hfill 
  \includegraphics[width=0.48\textwidth]{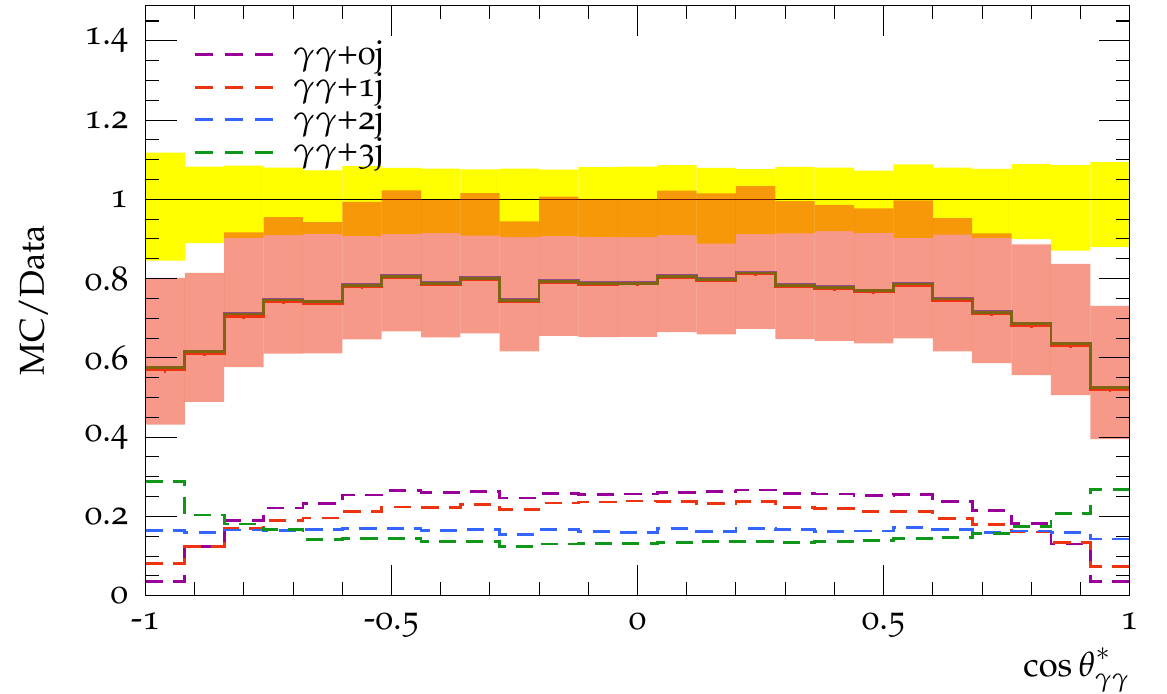} 
  
  \caption{MEPS@LO predictions for diphoton production at the 7 TeV LHC for the invariant mass (top left), transverse momentum (top right) and azimuthal separation (bottom left) of the photon pair and the polar angle of the harder photon in the Collins-Soper frame (bottom right). Data from the ATLAS experiment~\cite{Aad:2012aa} is represented by the black markers and the yellow uncertainty band. The solid red line represents the central prediction and the red markers (band) show the corresponding statistical (systematic) uncertainties. The solid blue and green line show predictions using different PDF sets as described in the main text.
The dashed coloured lines in the ratio plot demonstrate the composition of the central prediction from different jet multiplicities at the parton level.}
  \label{fig:loresults}
\end{figure}

\begin{figure}[tbp]
  \centering
  \includegraphics[width=0.48\textwidth]{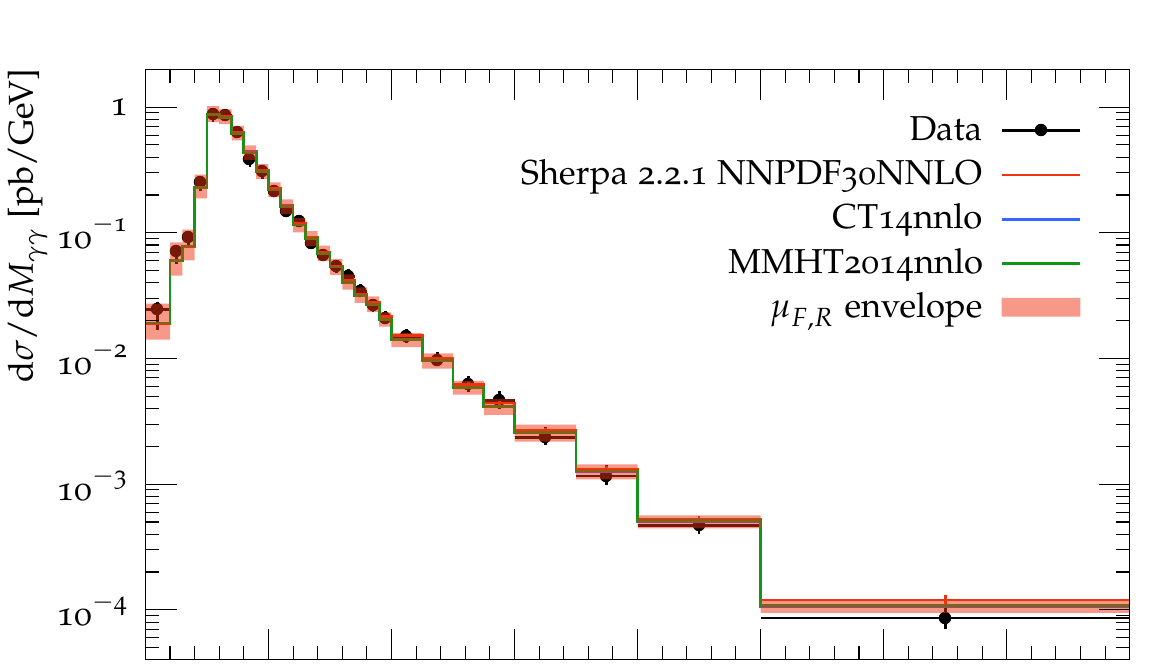}\hfill 
  \includegraphics[width=0.48\textwidth]{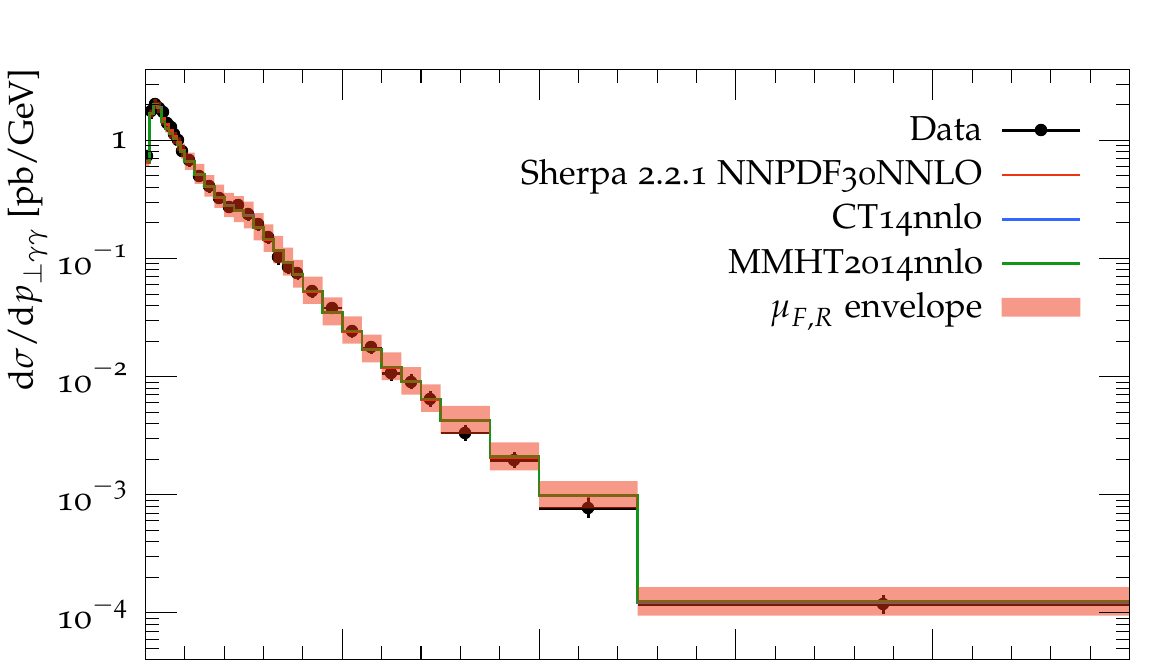}\\ 
  \includegraphics[width=0.48\textwidth]{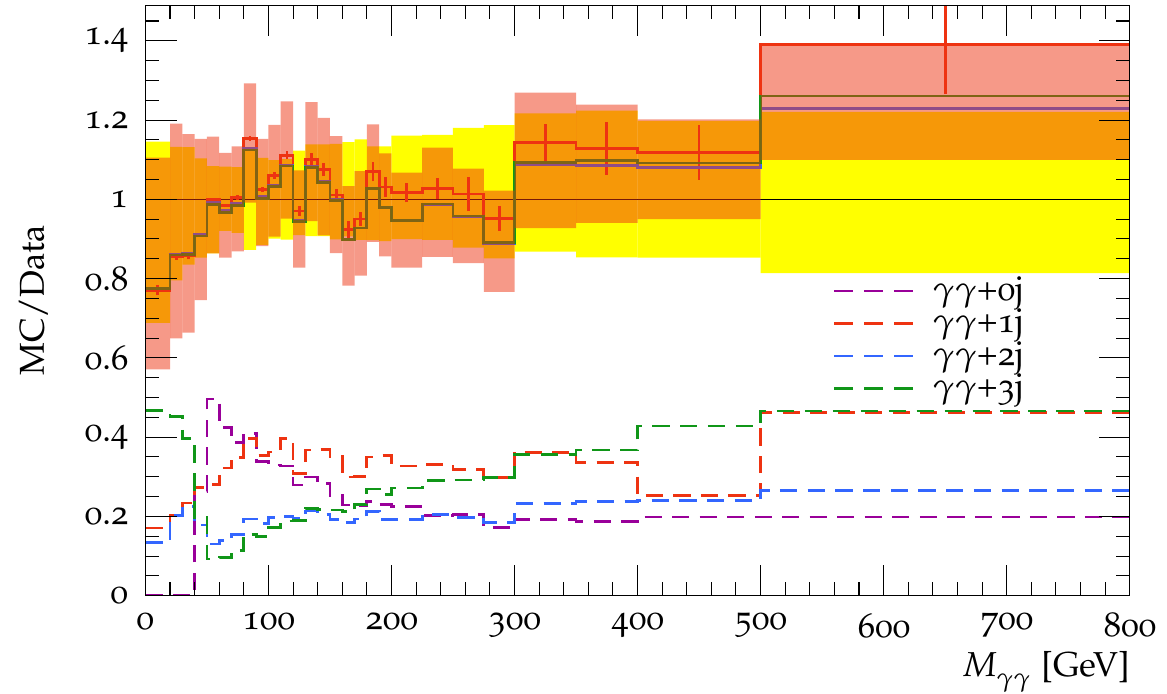}\hfill 
  \includegraphics[width=0.48\textwidth]{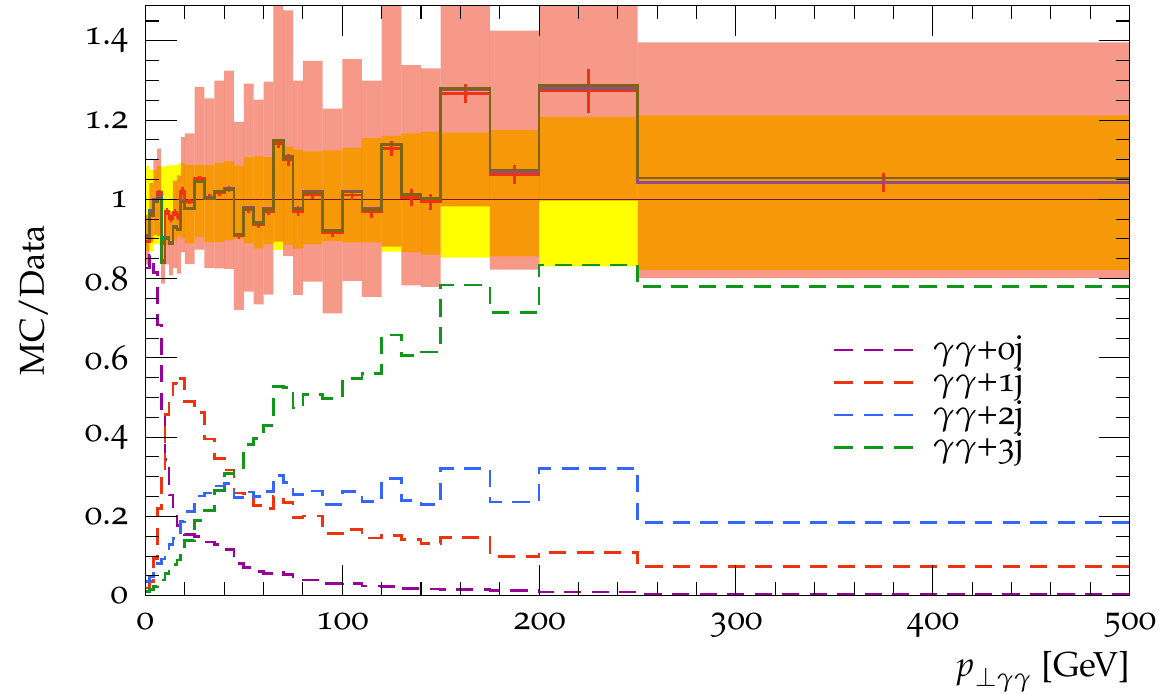}\\ 
  \includegraphics[width=0.48\textwidth]{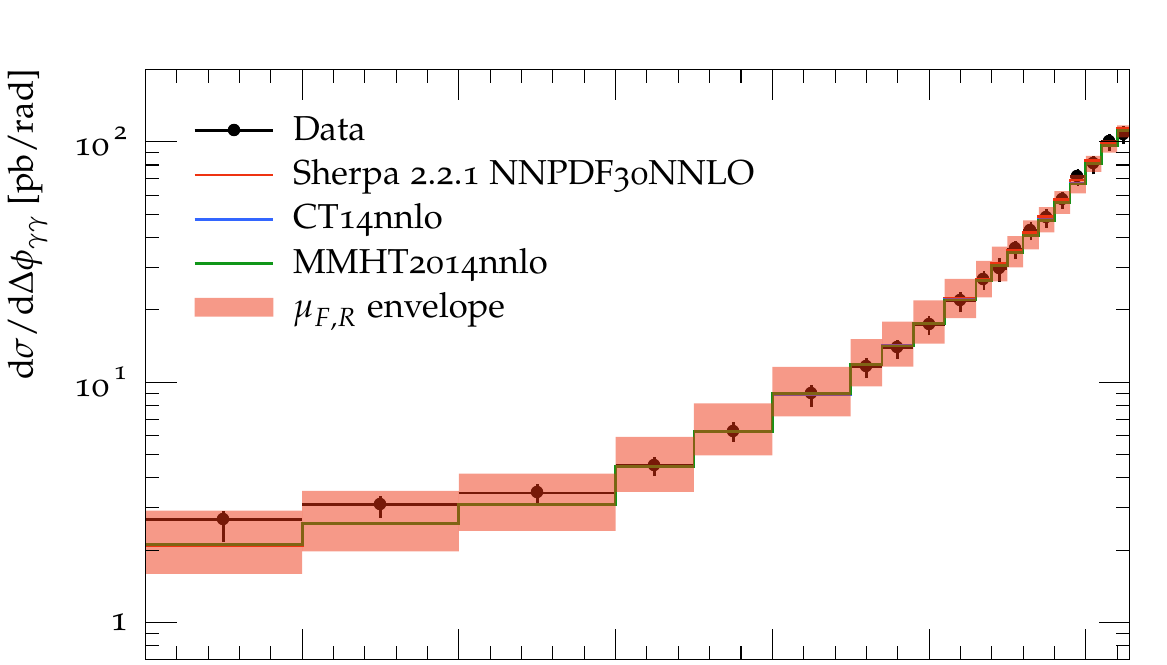}\hfill 
  \includegraphics[width=0.48\textwidth]{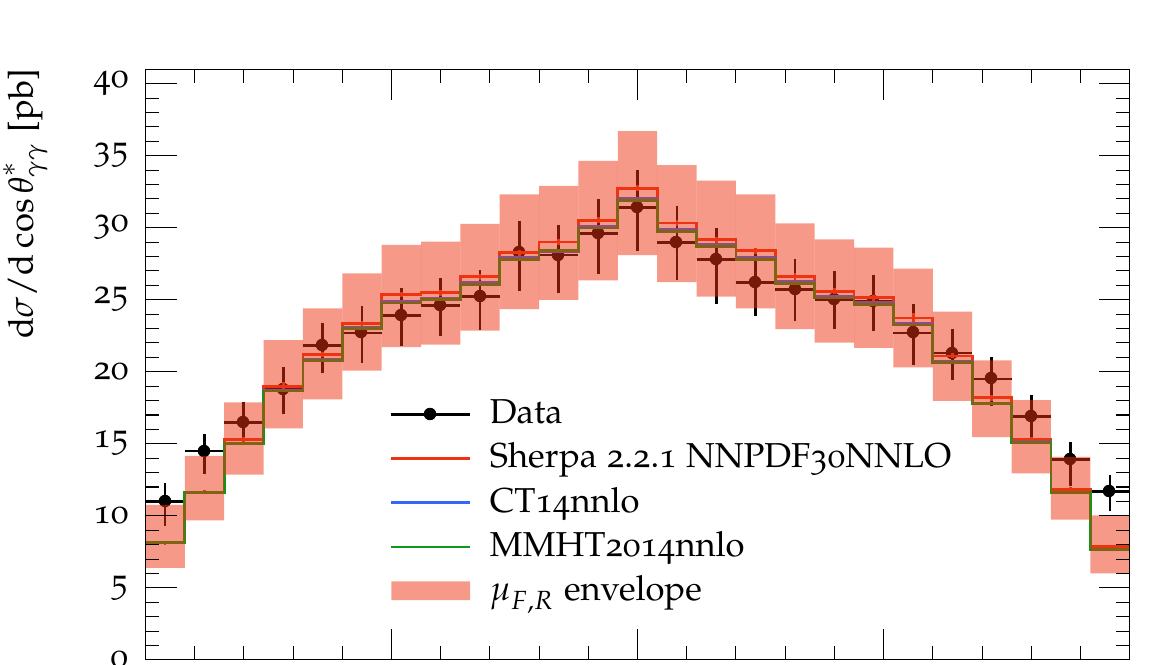}\\ 
  \includegraphics[width=0.48\textwidth]{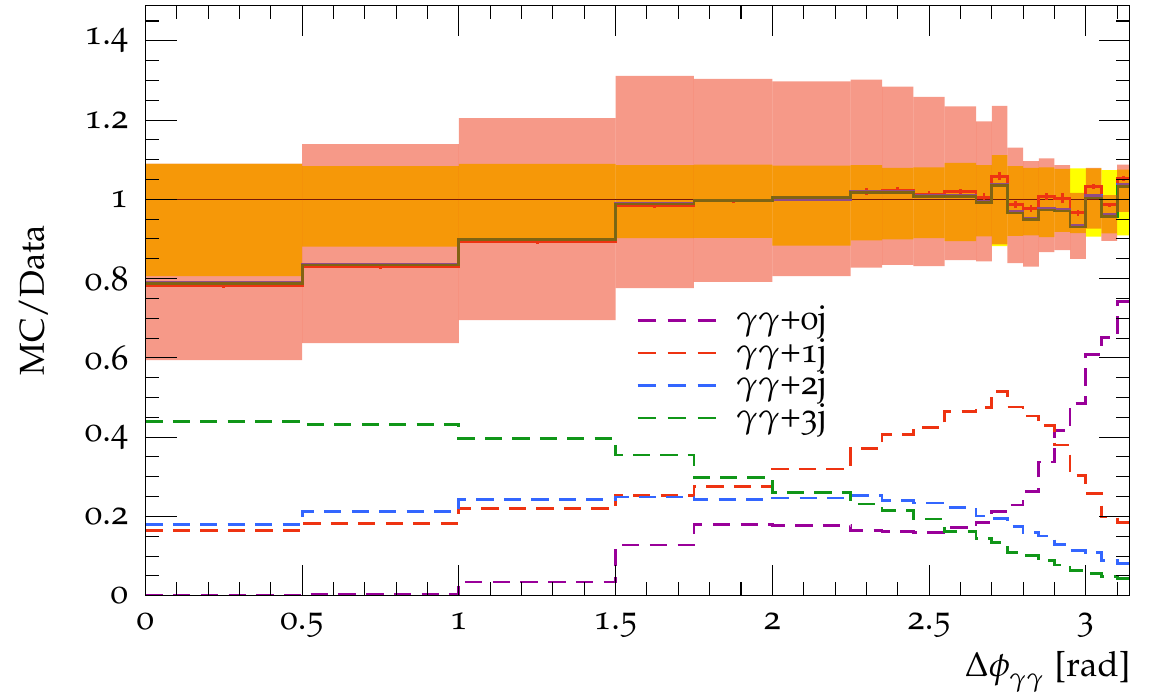}\hfill 
  \includegraphics[width=0.48\textwidth]{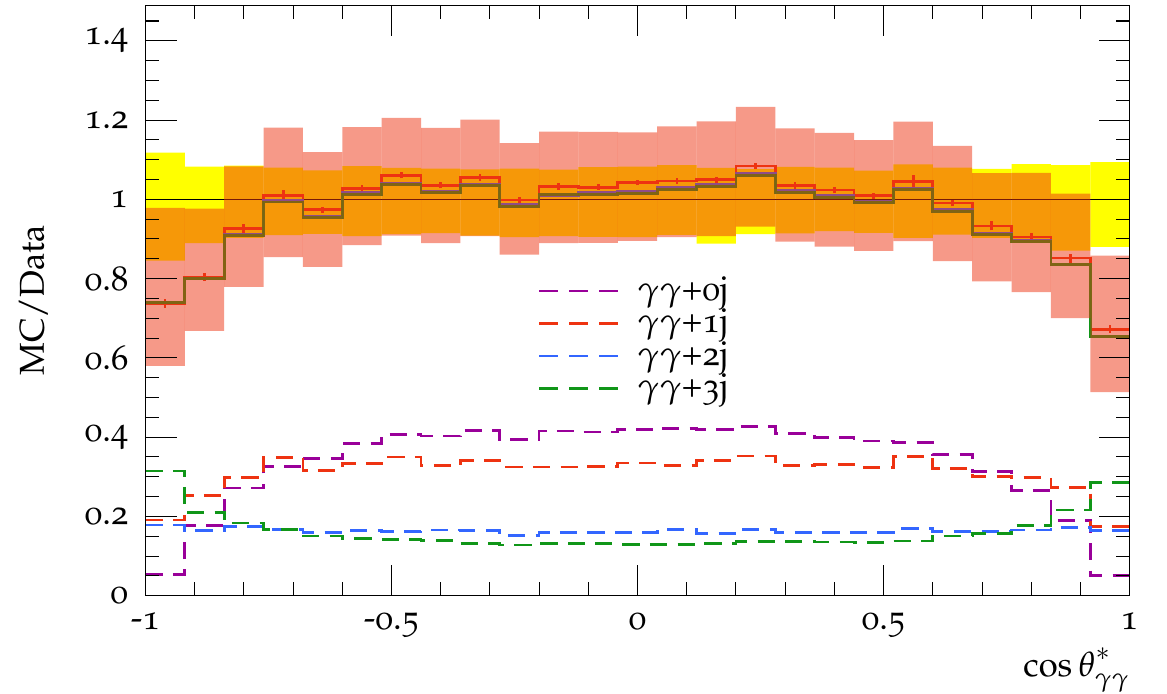} 
  
  \caption{MEPS@NLO predictions for diphoton production at the 7 TeV LHC for the invariant mass (top left), transverse momentum (top right) and azimuthal separation (bottom left) of the photon pair and the polar angle of the harder photon in the Collins-Soper frame (bottom right). Data from the ATLAS experiment~\cite{Aad:2012aa} is represented by the black markers and the yellow uncertainty band. The solid red line represents the central prediction and the red markers (band) show the corresponding statistical (systematic) uncertainties. The solid blue and green line show predictions using different PDF sets as described in the main text.
The dashed coloured lines in the ratio plot demonstrate the composition of the central prediction from different jet multiplicities at the parton level.}
  \label{fig:results}
\end{figure}

A more detailed picture emerges from the study of the differential
distributions displayed in Figs.~\ref{fig:loresults} and \ref{fig:results}.

The LO predictions show
strong shape deviations from data in the region of low transverse momentum
of the photon pair, yielding a deficit of up to $40\%$. This is strongly
improved in the NLO predictions, which agree well with data in this region
which is significantly affected by fragmentation photons.
Overall, the shape of the $p_{\perp\gamma\gamma}$ spectrum is described very
well throughout the whole range in the NLO prediction,
including the resummation region $p_{\perp\gamma\gamma}\to 0$, the hard region
$p_{\perp\gamma\gamma}>100\textrm{ GeV}$, and the intermediate region including
the well-known shoulder.

The azimuthal decorrelation of the two photons reveals larger shape differences
between the two predictions in the region $\frac{\pi}{2}<\Delta\phi<\pi$, with
the NLO prediction providing a shape more compatible with what is found in data.

The invariant mass distribution of the photon pair and the polar
angle of the harder photon in the Collins-Soper frame exhibit only minor
differences between the shapes of the two predictions.
The suppressed region below the peak in the $m_{\gamma\gamma}$ spectrum
and the region of $|\cos\theta^*|\to 1$ are slightly underestimated in both
cases.

Let us now turn to a discussion of the systematic uncertainties.
While they have been significantly reduced when considering the total cross
section, the uncertainties in specific regions of phase space still seem
fairly large when going from LO to NLO.

To understand this feature, it is instructive to first consider the
multi-jet process composition demonstrated by
the dashed lines in the ratio plots. Some regions like
$p_{\perp\gamma\gamma}>100\textrm{ GeV}$ and $\Delta\phi_{\gamma\gamma}\to 0$
are dominated by hard multi-jet emissions, i.e.\ the blue and green dashed
lines corresponding to 2 and 3 jets at the parton level, respectively.

These processes are simulated here only at leading order accuracy. Thus
they will lead to larger systematic uncertainties as seen in the increased size
of the scale variation band in these regions. This effect is amplified as one
is using a dynamical merging scale which can extend to very small values.
For a small value of the merging scale the cross section of the LO matrix
elements for $pp\to \gamma\gamma + 2,3$ jets is increased relative to the
NLO $pp\to \gamma\gamma + 0,1$ jets matrix elements. On the other hand, if
the merging scale would be chosen less inclusive, parts of the fragmentation
component would be missing in the sample. Furthermore, in that case the
additional emissions would be modelled by the parton shower, which would also
induce larger uncertainties that are not included in the bands displayed here.

On the other hand, in regions dominated by the NLO-accurate process simulation
one finds uncertainties comparable in size to the current experimental
uncertainties, thus demonstrating the improvements due to the application of
the MEPS@NLO merging approach.

\section{Conclusions}

A pedagogical overview of prompt photon production in the parton shower event
generator Sherpa was given. Modern generators like Sherpa improve the precision
of these predictions by including exact higher-order corrections in QCD
perturbation theory.
A particular difficulty in the context of prompt photon production lies
in the inclusive description of the fragmentation component.
The extent to which this is possible in different approaches has been discussed.

New predictions for the production of a prompt photon pair with one additional
jet at NLO QCD accuracy and matched and merged with lower and higher multi-jet
configurations were presented for the first time.
Good agreement with measurements from the ATLAS experiment was found and
the theoretical uncertainties have been studied.

The assessment of these uncertainties naturally leads to an outlook to future
work. To improve predictions in the regions dominated by multi-jet production
it is desirable to include NLO accurate matrix elements also for the higher
multiplicities, as far as computationally feasible. The still sizable global
uncertainty can be reduced only if a new scheme is devised that allows the
inclusion of NNLO-accurate calculations for the inclusive process in such a
merged sample. Furthermore, with additional work on the logarithmic accuracy
of the parton shower one would reduce a source of uncertainties which have not
been considered in this work and which might still play a large role in some
regions of the observables shown.

\section*{Acknowledgements}

We are grateful to Stefan H\"oche and Steffen Schumann for many fruitful
discussions on the subject and manuscript.
This research was supported by the German Research Foundation (DFG) under
grant No.\ SI 2009/1-1.
We thank the Center for Information Services and High Performance Computing
(ZIH) at TU Dresden for generous allocations of computing time.

\section*{References}

\bibliographystyle{amsunsrt_modp}
\bibliography{journal}

\end{document}